\documentclass[twocolumn]{aastex631}

\newcommand\dd{data--driven }
\definecolor{burgundy}{rgb}{0.502,0.0,0.125}

\received{}
\revised{}
\accepted{}
\submitjournal{ApJ}
\shorttitle{Hybrid data-driven simulations of eruption}
\shortauthors{Afanasyev et al.}
\graphicspath{{./}{figures/}}
\begin{document}
\title{Hybrid data-driven magnetofrictional and magnetohydrodynamic simulations of an eruptive solar active region}

\correspondingauthor{Andrey N. Afanasyev}
\email{andrei.afanasev@colorado.edu}

\author{Andrey N. Afanasyev}
\altaffiliation{DKIST Ambassador}
\affiliation{Laboratory for Atmospheric and Space Physics, University of Colorado Boulder \\
1234 Innovation Drive, Boulder, CO 80303, USA}
\affiliation{National Solar Observatory, University of Colorado Boulder, Boulder, CO, USA}
\affiliation{Institute of Solar-Terrestrial Physics of SB RAS, Irkutsk, Russia}

\author{Yuhong Fan}
\affiliation{High Altitude Observatory, National Center for Atmospheric Research, Boulder, CO, USA}

\author{Maria D. Kazachenko}
\affiliation{Laboratory for Atmospheric and Space Physics, University of Colorado Boulder \\
1234 Innovation Drive, Boulder, CO 80303, USA}
\affiliation{National Solar Observatory, University of Colorado Boulder, Boulder, CO, USA}
\affiliation{Department of Astrophysical and Planetary Sciences, University of Colorado Boulder, Boulder, CO, USA}

\author{Mark C.M. Cheung}
\affiliation{CSIRO, Space \& Astronomy, Epping, NSW, Australia}

\begin{abstract}
We present first results of the hybrid \dd magnetofrictional (MF) and data--constrained magnetohydrodynamic (MHD) simulations of solar active region NOAA~11158, which produced an X-class flare and coronal mass ejection on 2011 February 15. First, we apply the MF approach to build the coronal magnetic configuration corresponding to the SDO/HMI photospheric magnetograms by using the JSOC PDFI\_SS electric field inversions at the bottom boundary of the simulation domain. We then use the pre-eruptive MF state at about 1.5~hour before the observed X-class flare as the initial state for the MHD simulation, assuming a stratified polytropic solar corona. The MHD run shows that the initial magnetic configuration containing twisted magnetic fluxes and a 3D magnetic null point is out of equilibrium. We find the eruption of a complex magnetic structure consisting of two magnetic flux ropes, as well as the development of flare ribbons, with their morphology being in good agreement with observations. We conclude that the combination of the \dd MF and data--constrained MHD simulations is a useful practical tool for understanding the 3D magnetic structures of real solar ARs that are unobservable otherwise.
\end{abstract}


\section{Introduction} \label{sec:intro}
The understanding of the detailed mechanisms and triggers of solar eruptions is a major problem of solar physics. Although significant progress has been made in understanding the eruption precursor structures \citep[e.g., reviews by][]{Cheung-Isobe-2014LRSP...11....3C, Patsourakos-ISSIreview}, and many numerical models using idealised constructions are able to capture the eruption initiation and reveal the trigger mechanisms (e.g. \citealp{Antiochos_1999, Fan_2004, Torok-Kliem-2005ApJ...630L..97T, Mackay_2006, Karpen_2012, Jiang-2021NatAs...5.1126J, Hassanin_2022}), it remains challenging to model the realistic complex magnetic field evolution of the observed eruptive events. The models often use simplified magnetic configurations imitating the observations (the so called data-inspired models), or use the observational constrains at some point in time to build data-constrained (quasi-) static equilibria (e.g. the nonlinear force-free field (NLFFF) extrapolations) as the initial state (see \citealp{Chintzoglou_2019} for details of the model classification and examples). By using the time-evolving observational information during the course of the simulated evolution (e.g. photospheric vector magnetic fields), the simulation can better match the morphology of solar eruptions and provide the required quantitative agreement between models and observations, which is a subject of \dd simulations.

The methodology of \dd simulations is under active investigation (\citealp[see][for a review]{Jiang-rev-2022Innov...300236J}, and \citealp[][for comparison of several \dd methods]{Toriumi_2020}). Currently, there are several approaches to incorporate observational data into numerical simulations, for instances, based on the information on the lower boundary plasma velocities \citep{Hayashi_2019ApJ...871L..28H, Kaneko_2021ApJ...909..155K} or the electric fields inferred from the observed photosphere magnetograms (\citealp{Hayashi_2018ApJ...855...11H, Hoeksema_2020, Fan2022, Linton2022}). Another difficulty is that \dd magnetohydrodynamic (MHD) simulations are computationally expensive. Therefore, various simplifications to the governing set of the MHD equations are considered under the assumption of the dominating nature of the magnetic field, e.g. the magnetofrictional (MF) approach \citep{Cheung_DeRosa_2012, Pomoell_2019SoPh..294...41P, Hoeksema_2020, Lumme_2022A&A...658A.200L} or the magnetic relaxation in the zero-$\beta$ approximation \citep{Inoue_2014ApJ...788..182I, Inoue_2015ApJ...803...73I}.

In this study, we perform numerical simulations of the evolution of an active region by using a combination of the \dd MF and data-constrained MHD approaches.
As a target we choose active region (AR) NOAA~11158 (AR~11158, henceforth) that produced the first in the 24-th solar activity cycle X--class flare and a coronal mass ejection \citep[e.g.][]{Schrijver_2011, Petrie-2016SoPh..291..791P}. This AR has been intensively studied, both observationally and with numerical simulations \citep[e.g.][]{Sun_2012a, Sun_2012b, Cheung_DeRosa_2012, Inoue_2014ApJ...788..182I, Inoue_2015ApJ...803...73I, Fisher_2015, Zhao_2014, Kazachenko_2015, Kazachenko_2017, Hayashi_2018ApJ...855...11H, Hayashi_2019ApJ...871L..28H, Hoeksema_2020}.

Previously, a similar idea of a combined simulation approach was employed by \citet{Amari2014, Muhamad_2017}, who used the NLFFF extrapolation as the initial magnetic configuration for MHD simulations and applied various boundary conditions to initiate the eruption process. In addition, \citet{Muhamad_2017} analysed the structure of the reconnected field lines and compared that to the observed flare ribbons to detect the eruption trigger structure. \citet{Inoue_2014ApJ...788..182I, Inoue_2015ApJ...803...73I, Inoue_2018NatCo...9..174I} also chose an initial NLFFF configuration and found in the MHD part of their simulations of the AR~11158 evolution that the reconnection-caused movement of the footpoints of the highly-twisted field lines appropriately mapped the distribution of the observed two-ribbon flares. Very recently, \citet{Wang_Jiang_2022arXiv220808957W} performed the data-constrained MHD simulation by using the potential magnetic field as the initial condition and applying continuous sunspot rotation at the bottom boundary.
In another recent study, \citet{Liu_2022} carried out a data-constrained MHD simulation of the CME produced in AR 11520, where the initial state is an approximately force-free magnetic field constructed with the flux rope insertion method and MF relaxation (e.g. \citealp{vanBallegooijen2004ApJ...612..519V, Liu2018ApJ...868...59L}). They found that the initial force-free magnetic flux rope was already unstable with respect to the torus instability and immediately erupted in the MHD simulation without any further driving.

In this Paper, we report our first findings obtained from applying the hybrid simulation approach, \dd MF and data-constrained MHD, to the X-class flare eruption of AR~11158. To our knowledge, the results of such a hybrid approach are presented for the first time. In Section~\ref{sec:method}, we give a brief discussion of the numerical tools we used and describe the details of the simulation setups. Section~\ref{sec:results} presents the simulation results and discussion. We summarise our findings and discuss the follow-up work in Section~\ref{sec:concluions}.

\section{Methods} \label{sec:method}

To model the coronal magnetic field evolution of AR~11158, we perform hybrid simulations combining the \dd MF and data--constrained MHD simulations. First, we use the spherical MF method, in particular, its implementation within the Coronal Global Evolutionary Model (CGEM) framework \citep{Fisher_2015, Hoeksema_2020}. Originally being proposed as a method to obtain the NLFFF in the solar corona \citep{Yang-1986ApJ...309..383Y}, the MF method was extended to model the time-dependent quasi-equilibrium evolution of the coronal magnetic configuration (see, e.g. \citealp{Cheung_DeRosa_2012, Pomoell_2019SoPh..294...41P, Lumme_2022A&A...658A.200L} and references therein) by incorporating the photospheric electric fields inverted from the SDO/HMI magnetic observations as the lower boundary driving conditions \citep{Hoeksema_2020}. Unlike the standard MHD approach, the MF method evolves the magnetic field with the induction equation only, assuming the plasma velocity to be proportional to the Lorentz force. Due to the continuous data--based driving of the bottom boundary, the simulated magnetic configuration is able to quasi-statically build up the free magnetic energy associated with the coronal electric currents in the simulation domain. The MF method includes neither appropriate thermodynamics of the coronal plasma nor the real velocity field inside the simulation domain. However, being significantly less computationally expensive, it allows one to build up the pre--eruptive magnetic structures in the corona, as well as perform boundary data–driving spanning realistic (long) time scales.

Using the data-driven CGEM/MF model, we simulate the evolution of AR~11158 from the first appearance of the emerging magnetic flux on the photosphere around $t_0=$~14:10~UT on 2011 February 10 to the X-class flare around $t_{flare}=$~01:45~UT on 2011 February 15. For the sake of the subsequent MHD simulation, we use two times coarsened resolution of the available SDO/HMI data. The CGEM/MF code automatically calls the SDO/HMI Joint Science Operations Center (JSOC) pipeline software to retrieve the photospheric electric fields and then downsizes those to the halved resolution, $ds\approx730$~km. The uniform staggered simulation grid on a spherical wedge domain has 348, 348, and 174 cells for the azimuth angle $\varphi$, polar angle $\theta$, and radial coordinate $r$, respectively. The bottom boundary of the simulation domain corresponds to the JSOC SHARP for AR~11158, with the longitude ranging from $-10.4^{\circ}$ to $10.5^{\circ}$, and the latitude from $-30.8^{\circ}$ to $-9.9^{\circ}$. The radial domain size ranges from 1 to 1.36 solar radii ($R_{\odot}$).

The initial coronal magnetic field at $t_0=$~14:10~UT on 2011 February 10 in the CGEM/MF method is obtained using the ``nudging'' electric fields as described in \citet{Fisher_2020ApJS..248....2F} and starting with the zero--field assumption. In such an approach, the coronal magnetic field relaxes towards the force--free magnetic field matching the first available radial-field magnetogram, according to the original idea of the MF method. The same concept of ``nudging'' electric fields is also used during the whole \dd MF stage of our hybrid simulations to match the observed radial magnetograms. The top and side boundary conditions assume zero tangential components for the magnetic field vector, and zero--gradient conditions for the MF velocity vector. For more details of the MF method, we refer the reader to Section~4 in \citet{Hoeksema_2020}.

\begin{figure*}
\begin{interactive}{animation}{mov_B_cgem_hmi.mp4}
\fig{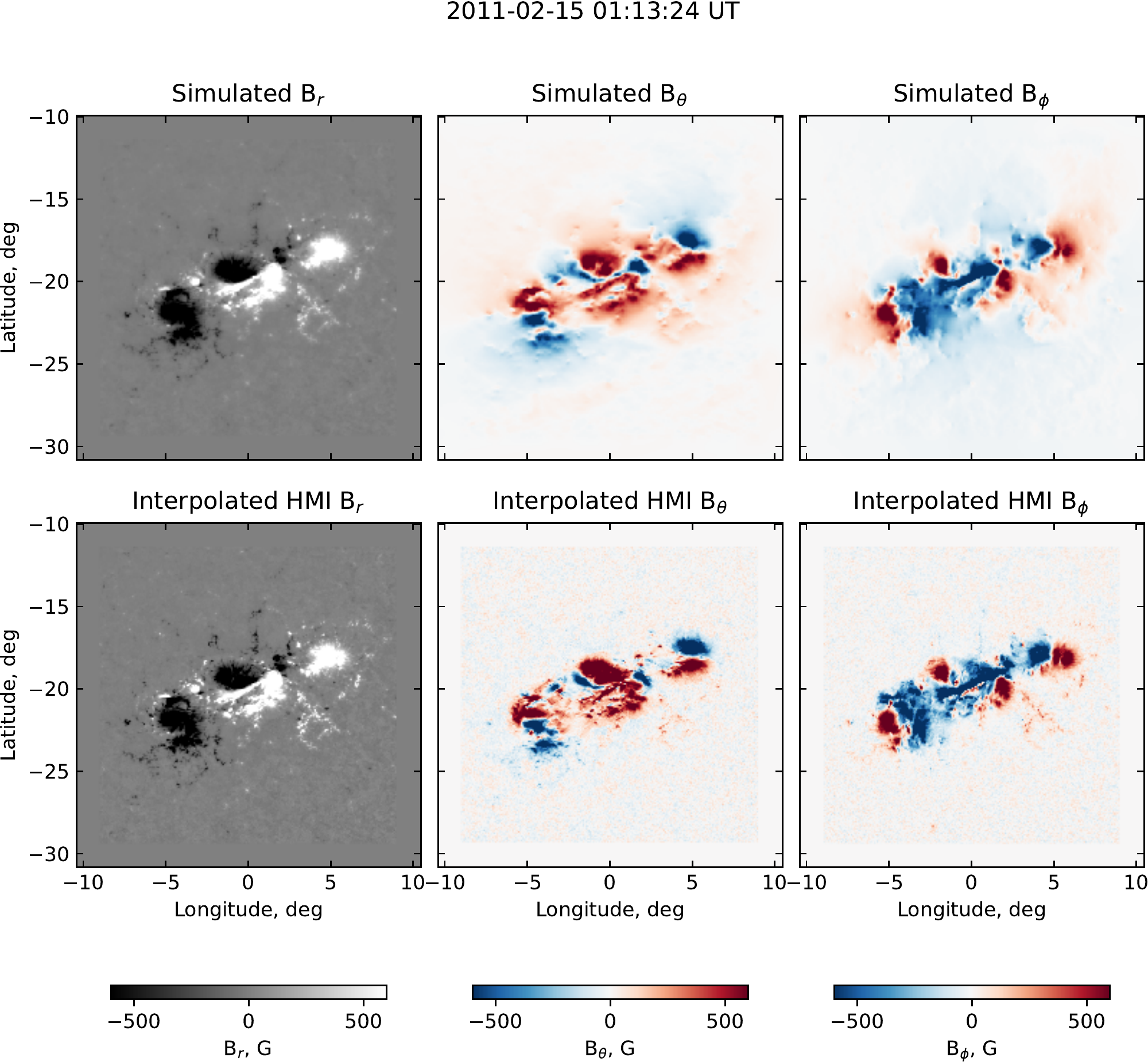}{0.89\textwidth}{}
\end{interactive}
\caption{Maps of the CGEM/MF simulated magnetic fields (top panels) at the bottom domain boundary and the SDO/HMI magnetograms (bottom panels) at 01:13~UT on 2011 February 15. This figure is sampled from an animation showing the evolution of the magnetic fields presented in the figure during the whole simuilation course from 14:10~UT on 2011 February 10 to 19:55~UT on 2011 February 15. The animation demonstrates good agreement between the CGEM/MF simulated magnetic fields and the SDO/HMI magnetograms. See Section~\ref{sec:method} for details. (An animation of this figure is available.)
\label{fig:mf-hmi-comparison}}
\end{figure*}

To demonstrate the data-driving concept of the CGEM/MF model using the photospheric electric fields, we compare the simulated magnetic fields at the bottom domain boundary with the SDO/HMI observations. Figure~\ref{fig:mf-hmi-comparison} shows an example of such comparison for all three components of the magnetic field at $t=$~01:13~UT on 2011 February 15. We find very good agreement between the simulated and observed magnetic field components. All the large-scale structures of different polarities (off course, on a scale of a solar active region) are present both in the simulations (top panels) and the observations (bottom panels), although the small-scale differences can easily be detected. The bottom panels show the observed (not-downsized) magnetic field component linearly interpolated to the corresponding simulation output time.

At the next step, we use the pre-eruptive MF state at 00:07~UT on 2011 February 15, about 1.5 hour before the observed X2.2 flare, as an initial magnetic configuration for the MHD simulation. By taking an earlier state, we look to see if the magnetic configuration of choice would result in an immediate release of the stored magnetic stresses or rather slowly evolve towards that. However, we still choose a MF state to be close enough to the onset of the X-class flare because AR~11158 also showed other activity on 2011 February 14 (in particular, M-class flare at around 18:00~UT). We also want to minimize the length of time in the build-up phase for the MHD simulation because of its relatively higher numerical diffusion. By 00:07~UT we expect the magnetic configuration to be sufficiently energised by the bottom--boundary data--driving and lead to an eruption onset in the subsequent MHD simulation.

The MHD simulation is performed using the Magnetic Flux Eruption (MFE) code by \citet{Fan_2017}. The governing equations and the boundary conditions used for this simulation are the same as those in \citet{Liu_2022}. We solve the set of the MHD equations, where the momentum equation includes the semi-relativistic correction to handle the high values of the Alfv{\'e}n speed in the AR. As described in \citet{Liu_2022}, the internal energy equation includes explicitly the non-adiabatic effects of the field-aligned thermal conduction and the heating due to numerical dissipation of the magnetic and kinetic energies, while neither optically thin radiative cooling nor any empirical corona heating terms are included. An ideal gas with the reduced adiabatic index $\gamma = 1.1$ is assumed to maintain the high coronal plasma temperature in the simulation domain.

In the MFE setup, we use the same staggered spherical grid as in the MF part of our hybrid simulations and assume the following polytropic hydrostatic solar corona stratified over $r$
\begin{eqnarray}
    \rho = \rho_{\odot} \left[ 1 - \frac{G M_{\odot}}{R_{\odot}} \frac{\rho_{\odot}}{p_{\odot}}
    \left( 1 - \frac{1}{\gamma} \right)
    \left( 1 - \frac{R_{\odot}}{r} \right)
    \right] ^{\frac{1}{\gamma - 1}},       \label{eq:polytropic-rho}       \\
     p = p_{\odot} \left[ 1 - \frac{G M_{\odot}}{R_{\odot}} \frac{\rho_{\odot}}{p_{\odot}}
    \left( 1 - \frac{1}{\gamma} \right)
    \left( 1 - \frac{R_{\odot}}{r} \right)
    \right] ^{\frac{\gamma}{\gamma - 1}},  \label{eq:polytropic-pres}
\end{eqnarray}
where $\rho_{\odot} = \rho(R_{\odot})$ and $p_{\odot} = p(R_{\odot})$ are the plasma density and pressure values at the corona base, respectively, that correspond to the plasma temperature of $1.6\times 10^{6}$~K and the proton number density of $1.0 \times 10^{9}$~cm$^{-3}$, $G$ and $M_{\odot}$ are the gravitational constant and solar mass, respectively. The hydrostatic stability of the atmosphere described by Equation~\ref{eq:polytropic-rho} and \ref{eq:polytropic-pres} with the chosen model parameters has been confirmed by a separate simulation run with zero magnetic field. To normalise the code equations, the following units are used: the length unit $\hat{L}=R_{\odot}$, the magnetic field unit $\hat{B} = 20$~G, the density unit $\hat{\rho} = 8.365 \cdot 10^{-16}$~g~cm$^{-3}$, and the temperature unit $\hat{T} = 1.0$~MK.

The boundary conditions at the top and side boundaries include the zero--gradient extrapolations for the plasma density, internal energy, and velocity components with the additional no--inflow conditions for the normal to the boundaries velocity component, as well as the extrapolations of the electric field values to provide the divergence-free boundaries for the magnetic field components. At the bottom boundary, we assume a line-tied boundary with zero velocity and zero electric field. For the thermodynamic lower boundary conditions, the temperature at the first grid zone above the lower boundary surface is fixed to its initial value, but the density (or pressure) is time varying, being proportional to the temperature in the grid zone above, so as to provide a variable coronal base pressure that increases with the temperature gradient and hence the downward heat conduction flux.

The MHD simulation uses the initial magnetic configuration obtained from a snapshot of the \dd MF simulation and the fixed line-tied lower boundary conditions to model the subsequent dynamic evolution without further boundary driving, and therefore falls into the so--called data--constrained class of simulations. Thus, we apply a hybrid simulation approach to study AR~11158.

\begin{figure*}
\fig{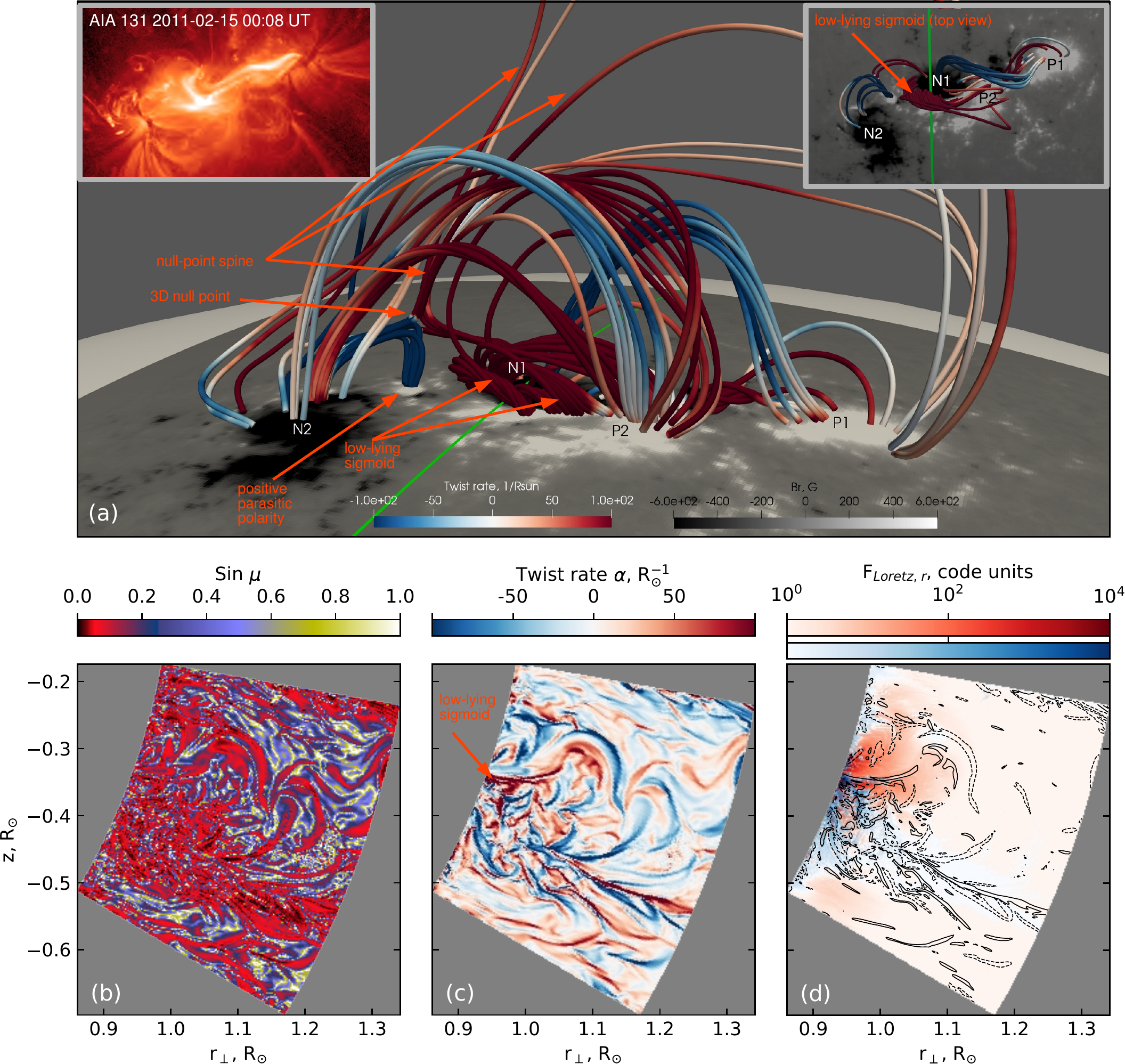}{0.99\textwidth}{}
\caption{Pre-eruptive MF magnetic configuration at $t_1=$~00:07~UT on 2011 February 15, about 1.5 hour before the X2.2 flare. The top panel (a) shows the 3D structure of magnetic field lines coloured by their twist per unit length, with the bottom background showing the radial component of the magnetic field. The right inset figure in panel (a) shows the top view of the field lines. The left inset figure in panel (a) shows the SDO/AIA~{131\,\AA} context image at 00:08~UT on 2011 February 15. The pairs of the magnetic polarities are marked as N1--P1 and N2--P2. The bottom panels show the vertical meridional slices of the simulation domain at $\varphi = -1.17^{\circ}$ with maps of the angle between the magnetic field and electric current (panel b), twist rate (panel c), and positive (red scale) and negative (blue scale) radial component of the Lorentz force (panel d) with the superimposed twist rate contours of $\pm 60 \, R_{\odot}^{-1}$. The green line in panel (a) corresponds to the base of the vertical meridional slice used in the bottom panels. See Section~\ref{sec:results} for details.
\label{fig:mf-state}}
\end{figure*}

\section{Results and Discussion} \label{sec:results}

Figure~\ref{fig:mf-state} presents the properties of the pre-eruptive magnetic field from the MF approach at $t_1=$~00:07~UT on 2011 February 15, around 1.5 hour before the observed X2.2 flare. The total magnetic energy in the volume, $E_m\approx1.2 \times 10^{33}$~erg, is in agreement with the results obtained with the NLFFF extrapolation by \citet{Sun_2012b}. Figure~\ref{fig:mf-state}a shows the three-dimensional (3D) structure of magnetic field lines of the pre-eruptive configuration. The colour of the field lines shows values of the twist rate, or magnetic twist per unit length, along them, $\alpha = \mathbf{j} \cdot \mathbf{B} /B^2$. The bottom boundary of the domain shows the radial field magnetogram. We denote pairs of the magnetic polarities as N1--P1 and N2--P2, using notation of the colliding magnetic bipoles from \citet[][]{Chintzoglou_2019}. We find a low-lying sigmoid-shaped sheared magnetic flux above the collisional PIL between polarities N1--P2, which contains field lines with high positive twist rates and corresponds in its shape to the observed Extreme-Ultraviolet (EUV) sigmoid in AR~11158 (see the left inset figure in panel~\ref{fig:mf-state}a) showing the observed SDO/AIA~{131\,\AA} image of AR~11158. That low-lying sigmoid can be potentially a magnetic flux rope (MFR) with the magnetic twist number more than unity or become such during the MHD evolution of the initial MF magnetic configuration. In particular, \citet{Chintzoglou_2019} referred to that sigmoidal structure as a MFR. We also find both positive-twisted and negative-twisted fluxes higher up in the corona. The east part of the magnetic configuration includes a small parasitic positive polarity between N1 and N2 that is connected to a 3D magnetic null point whose spine overlies the low-lying sigmoid and the other higher-lying twisted fluxes and extends to polarities P1 and P2.

Figure~\ref{fig:mf-state}b shows a map of the sine of the angle between the magnetic field and electric current, $\sin \mu = \mathbf{j} \times \mathbf{B} / \left( j B \right)$, in the vertical meridional slice of the simulation domain at $\varphi = -1.17^{\circ}$. The position of the slice is shown as its intersection with the bottom boundary of the simulation domain, marked by the green line in Figure~\ref{fig:mf-state}a. The magnetic field configuration is apparently not force-free, although it is quite close to the force-free state (with $\sin \mu = 0$) lower in the simulation domain and at the regions of a high twist rate (see also Figure~\ref{fig:mf-state}c). The current-weighted mean value of $\sin \mu$ for the domain (as calculated in \citealp{Toriumi_2020}) is 0.19, or 11$^{\circ}$ for the angle $\mu$ itself. We also find that the spatial distribution of the force-freeness is highly non-uniform, with small-scale fluctuations produced by the small-scale variations in the lower boundary magnetic field.

Figure~\ref{fig:mf-state}c shows the distribution of the twist rate in the same meridional slice marked by the green colour. The small-scale region of a high positive twist rate close to the bottom boundary (seen as the saturated red colour and marked by the arrow) corresponds to the low-lying sigmoid found in Figure~\ref{fig:mf-state}a. The other relatively high-twisted regions correspond to the above-lying magnetic fluxes. In particular, to construct the field lines in Figure~\ref{fig:mf-state}a, we use seed points at the regions of high positive and negative twist in that meridional slice.

Figure~\ref{fig:mf-state}d shows the radial component of the Lorentz force (per unit volume) in the same meridional slice marked by the green colour. To present its magnitude in the logarithmic scale, we plot its absolute value, with the red colour showing its positive magnitude and the blue colour showing the negative one. Note that we clip the colour bars at $10^4$~code units (with the force code unit of $5.7 \times 10^{-4}$~G$^2$~km$^{-1}$), so the values very close to the bottom are slightly saturated. In addition, we superimpose contours of the constant twist rate ($\pm 60 \, R_{\odot}^{-1}$) to see how the Lorentz force distributes over the twisted flux regions, with the solid (dashed) contours for the positive (negative) twist rate values. From this panel, we find that the initial magnetic field has a significant net upward Lorentz force in the lower part of the domain. However, as the contours show, the low-lying sigmoid field contains both upward and downward Lorentz forces. We also note that the large region to the north from the sigmoid has the dominating outward-directed Lorentz force. The non-equilibrium initial magnetic field from the MF calculation leads to the subsequent dynamic evolution in the MHD simulation as described below.

\begin{figure*}
\fig{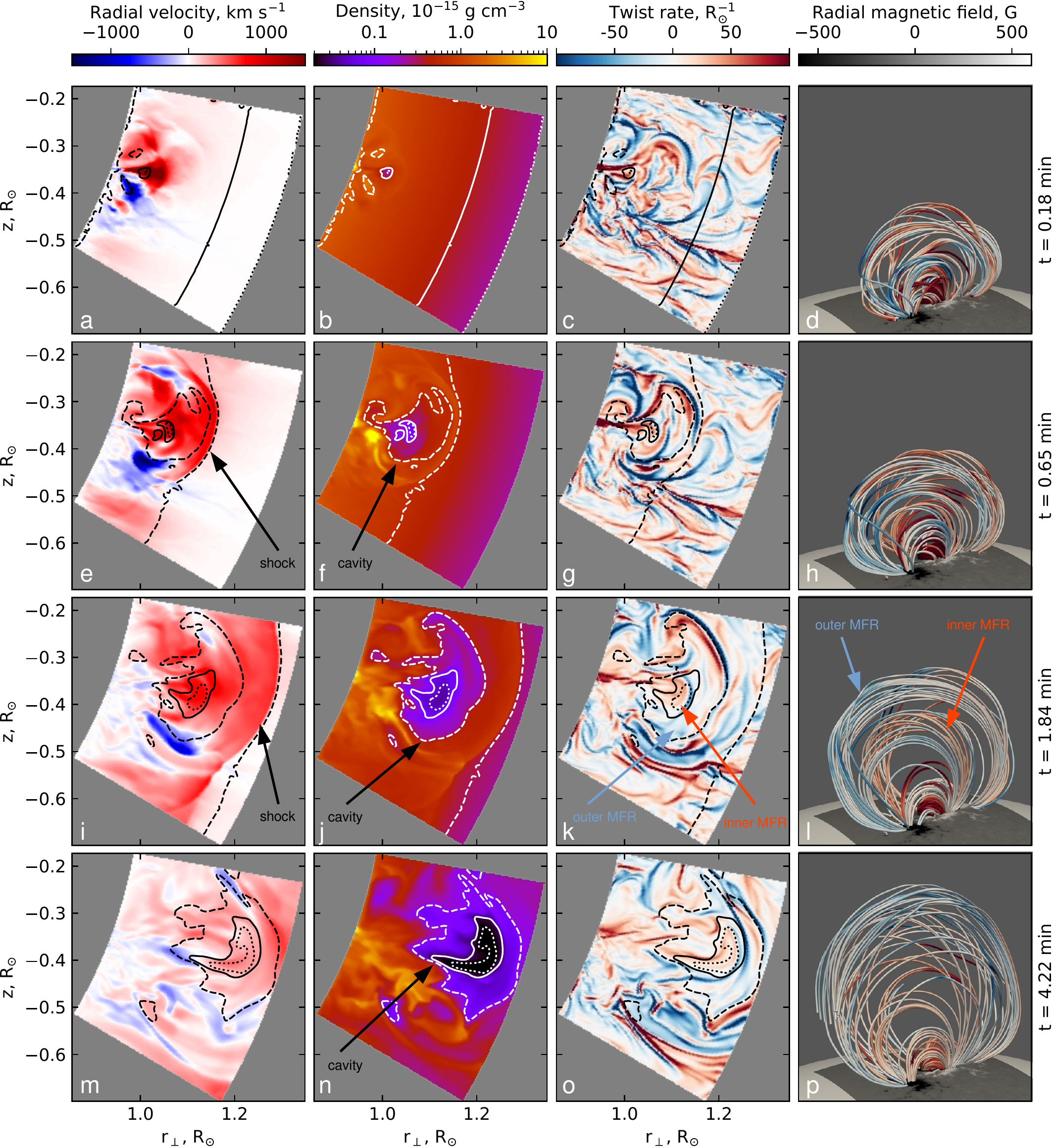}{1.0\textwidth}{}
\caption{Evolution of the radial velocity, plasma density, and twist rate in the meridional slice at $\varphi = -1.17^{\circ}$, and the structure of magnetic field lines at different times in the MHD run of the pre-eruptive MF configuration. The dotted, solid, and dashed contours are plotted using the density maps and superimposed to the other maps, and correspond to the value of [1.2, 2.0, 8.0] times the minimum density value at each snapshot, respectively. An animation of this figure is available. The animation shows the MHD evolution presented in this figure during the time period of 5.95~min starting from the MHD run onset. The contour annotations are the same as in the figure. See Section~\ref{sec:results} for details.
\label{fig:evolution}}
\end{figure*}

Figure~\ref{fig:evolution} shows the MHD evolution of AR~11158 after $t_1=$~00:07~UT on 2011 February 15. From left to right, each column shows maps of the radial plasma velocity, plasma density, and twist rate in the meridional slice at $\varphi = -1.17^{\circ}$, as well as the 3D structure of the magnetic field lines. We also draw the plasma density contours corresponding to the values of [1.2, 2.0, 8.0] times the minimum density value at each snapshot onto the maps of the radial velocity and twist rate. The initial MHD state includes the radially stratified polytropic atmosphere in a hydrostatic equilibrium and the non-force-free magnetic configuration presented in Figure~\ref{fig:mf-state} with the low-lying sigmoid and coronal null-point. The region surrounding the low-lying sigmoid field immediately accelerates outward (Figure~\ref{fig:evolution}a), developing a region of rapid outward eruption with a shock-like wavefront (annotated in Figure~\ref{fig:evolution}e,i). That erupting region has mostly positive values of the radial component of the Lorentz force (see Figure~\ref{fig:mf-state}d), which implies that the plasma is driven upwards by the initial Lorentz force due to the non-equilibrium initial magnetic field.

Simultaneously, a density cavity forms in the erupting region behind the shock and the dense sheath, as seen from the density contours in the two left columns. The cavity consists of several parts. The least dense inner part (enclosed within the dotted density contour) corresponds to the positively twisted flux (with mainly red colour in the twist rate maps in the third column). It originates from the region just above the low-lying sigmoid initially found above the collisional PIL. This least dense part of the cavity forms the inner MFR, which remains positively twisted, and it is surrounded by an outer cavity (enclosed by the dashed density contours) which contains a significant amount of negatively twisted flux. During the later stage of the simulation (see e.g. Figure~\ref{fig:evolution}l), the footpoints of the inner MFR are rooted in N2 and P2 polarities. The outer cavity forms an outer MFR with dominating negative twist rates as can be seen in the twist maps. During the later stage of the simulations, its footpoints are rooted in N2 and P1 polarities. The spatial structures of both the inner and outer MFRs can be seen in the 3D visualisation of the magnetic field lines (Figure~\ref{fig:evolution}, right column) during the later stage (in particular, at Figure~\ref{fig:evolution}l with the annotated MFRs). Note that here we define a set of twisted field lines as a MFR not from the total twist number along field lines but using the values of the magnetic field twist rate $\alpha$. The field lines in the right column snapshots are plotted by tracing field lines from a set of Lagrangian tracers selected in the cavity region as well as in the regions of the high twist rate (see Figure~\ref{fig:mf-state}c) and tracked in the velocity field.

\begin{figure*}
\fig{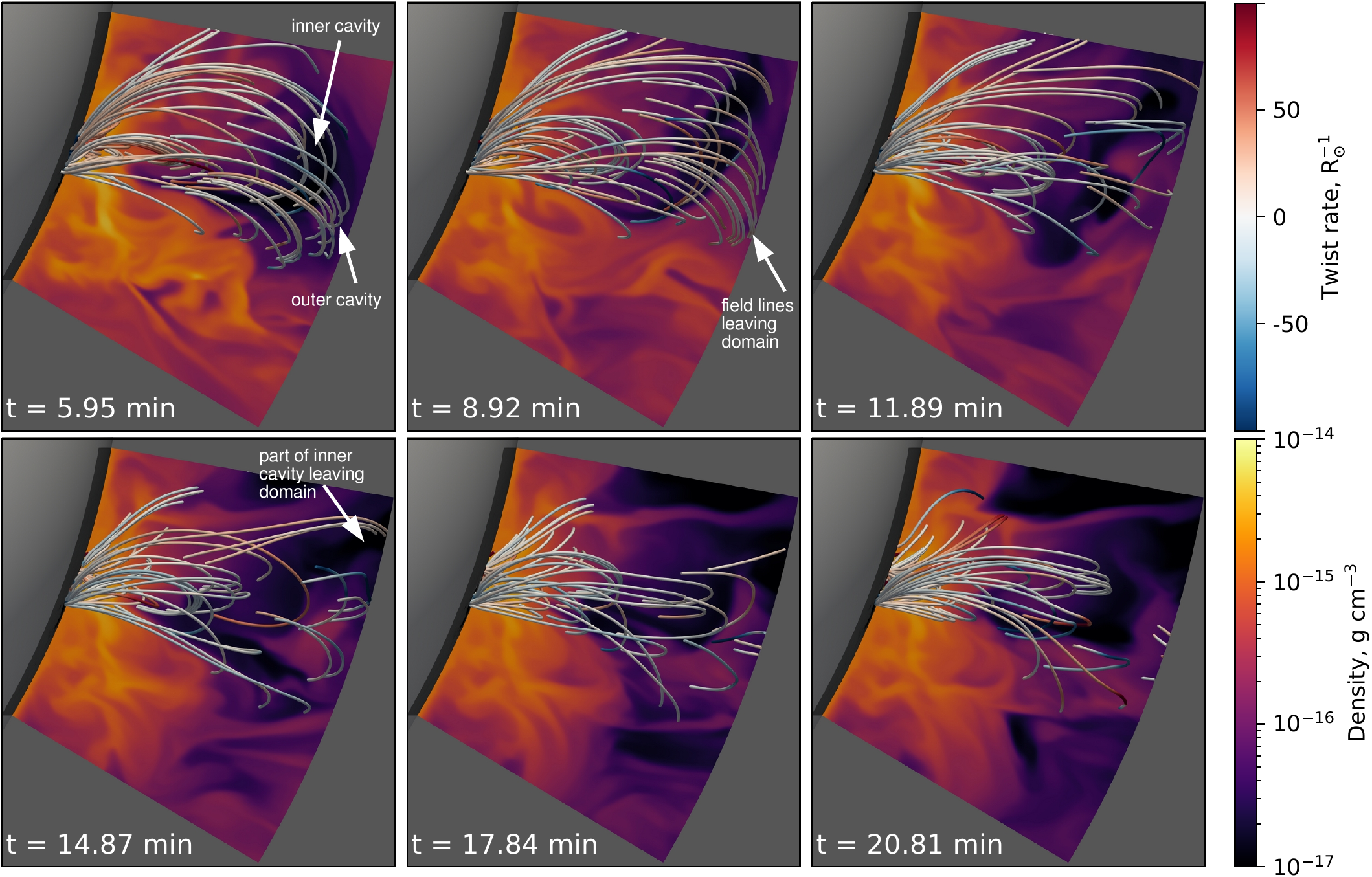}{0.99\textwidth}{}
\caption{Later stage of the eruption of the density cavity and related magnetic field lines in the MHD run as seen in the vertical meridional slice at $\varphi = -1.17^{\circ}$. An animation of this figure is available. The animation shows the evolution presented in this figure during the time period of 29.73~min starting from the MHD run onset. See Section~\ref{sec:results} for details.
\label{fig:slice-n-fieldlines}}
\end{figure*}

In addition, we analyse if the magnetic field associated with the erupting region is ejected out of the simulation domain. Figure~\ref{fig:slice-n-fieldlines} shows the later stage (6~minutes after the MHD run onset) of the density cavity evolution in the same meridional slice at $\varphi = -1.17^{\circ}$ with the tracked 3D magnetic field lines permeating the slice. Like the analysis of the twist rate evolution (see Figure~\ref{fig:evolution}), it confirms that the outer MFR corresponding to the outer density cavity leaves the simulation domain first, pushed by the inner MFR below it. The inner density cavity then also partially leaves the domain. Thus, the two MFRs of different twists, inner and outer, constitute the CME in this eruptive event in our MHD simulation.

\begin{figure*}
\fig{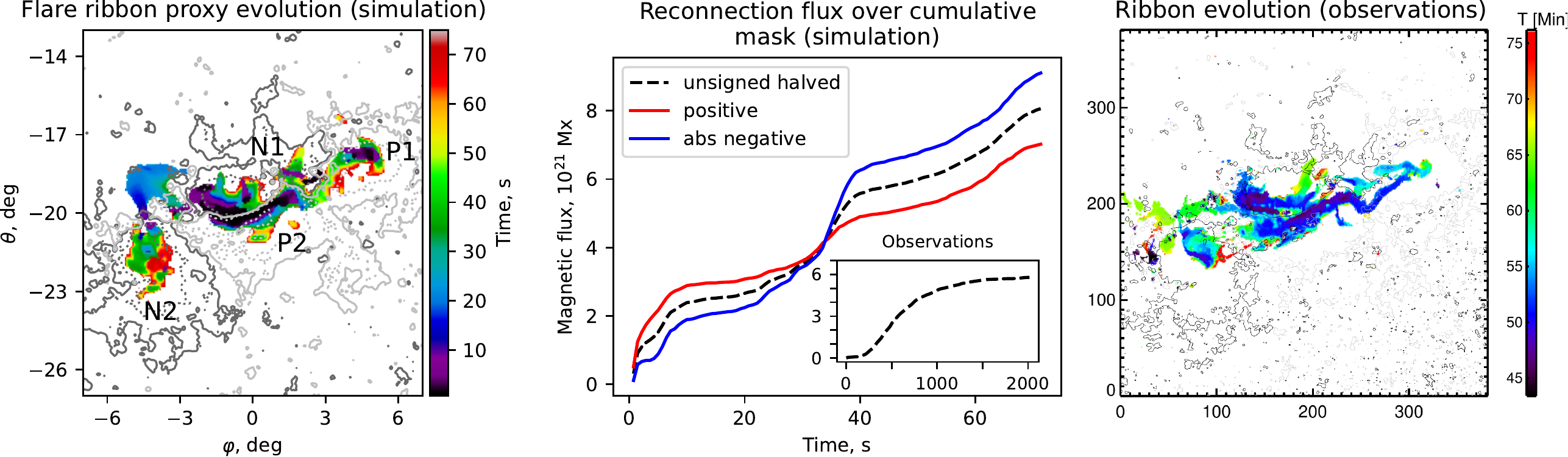}{0.99\textwidth}{}
\caption{Evolution of the plasma temperature enhancements near the bottom boundary of the simulation domain as a flare ribbon proxy (left panel) versus EUV SDO/AIA~{1600\,\AA} observations of flare ribbons after 01:00~UT on 2011 February 15 (right panel, adapted from \citealp{Kazachenko_2017}). In the left panel, the light(dark)-grey contours represent positive (negative) radial magnetic field values of 300~G (dotted lines) and 30~G (solid lines), and the magnetic polarities are marked as in Figure~\ref{fig:mf-state}. The middle panel shows the simulated and derived from observations reconnection flux over the cumulative mask. See Section~\ref{sec:results} for details.
\label{fig:ribbons}}
\end{figure*}

At the next step, we analyse the MHD part of the simulation to understand the structure and dynamics of flare ribbons. Flare ribbons are brightenings in $H_{\alpha}$ and {1600\,\AA} UV emission in the upper chromosphere (e.g. \citealp{Forbes_2000, Fletcher2011}). These brightenings are considered to be caused by the precipitation of the non-thermal particles accelerated by the magnetic reconnection. The flare ribbons show the footpoints of the reconnected field lines in the flare arcade. In our MHD simulation that includes the field-aligned thermal conduction and numerical resistive heating in the energy equation (see Section~\ref{sec:method}), the heat produced in the flare reconnection region propagates along the magnetic field lines towards the corona base via the field-aligned thermal conduction. This reconnection heating is very pronounced (up to about 100~MK), so the related temperature enhancements appear to exceed significantly the temperature variations due to the adiabatic plasma compressions. \citet{Cheung2019} also found such super-hot temperatures in their flare model. By detecting the temperature enhancements close to the bottom boundary of the simulation domain, we find the position of the flare ribbons. In what follows, we discuss simulated dynamics of flare ribbons by using the temperature enhancement near the lower boundary as ribbon proxy. We note that our approach to detect flare ribbons differs from existing approaches such as the calculation of the spatial variance of the field line connectivity \citep{Toriumi_2013ApJ...773..128T, Inoue_2014ApJ...788..182I} or the analysis of the rapid changes in the field line lengths \citep{Lynch_2019ApJ...880...97L, Lynch_2021}.

Figure~\ref{fig:ribbons} shows the evolution of the simulated flare ribbons derived from the footpoint temperature enhancements. The left panel shows their spatial and temporal variation. The colour represents the time elapsed since the MHD simulation start at $t_1=$~00:07~UT on 2011 February 15. The light-grey and dark-grey contours show positive and negative radial magnetic field values of $\pm 300$~G (dotted lines) and $\pm 30$~G (solid lines), respectively. To detect the spatial structure and location of the ribbons, we use a varying mask for temperature enhancements with a dynamic threshold value. In particular, for each simulation snapshot \textit{i}, we set a threshold value of $c_i=\kappa_{1}(\log_{10} T_{i,max}-\kappa_{2}\log_{10} T_{i,median})$, using the horizontal maps of the plasma temperature just above the bottom domain boundary, where $\log_{10} T_{i,max}$ and $\log_{10} T_{i,median}$ are the maximum and median logarithmed values for a snapshot, respectively, and $\kappa_1$ and $\kappa_2$ are constants determined empirically for the whole dataset. The reconnection flux presented in the middle panel is then calculated over the cumulative mask (see, e.g., \citealp{Kazachenko_2017}). The red and blue lines show the positive and negative reconnection fluxes, respectively, and the black dashed line shows the halved total unsigned reconnection flux. For comparison, in the inset panel we show the observed reconnection flux, and in the right panel we show the observed evolution of the flare ribbons as seen in EUV SDO/AIA~{1600\,\AA} images during the X-class flare in AR~11158 from \citet{Kazachenko_2017}. The colour bar in the right panel shows the time after 01:00~UT on 2011 February 15.

From Figure~\ref{fig:ribbons}, we find that the flare ribbons originate near the collisional PIL between N1 and P2 polarities, clearly showing that the initial magnetic reconnection is related to the initial low-lying sigmoid (see Figure~\ref{fig:mf-state}). Then the flare ribbons separate, moving away from the PIL. The positive south ribbon extends significantly along the PIL towards the west, while the north ribbon has a hook-shaped geometry, outlining the negative footpoints of the initial low-lying sigmoid in agreement with the 3D generalisation of the standard flare model \citep[see e.g.][]{Aulanier_Dudik_2019}. Simultaneously, we detect the ribbons in polarity P1 and in the west negative parasitic polarity (although this parasitic polarity could be considered as a part of polarity N1 as the radial magnetic field contours show) that are apparently associated with the positive-twisted low-lying flux seen on the right-hand side in Figure~\ref{fig:mf-state} between P1 and N1 polarities. However, these ribbons do not develop so significantly as their counterparts around the collisional PIL.

After about 15~seconds since the MHD run onset (which corresponds to the saturated blue colour at the left panel of Figure~\ref{fig:ribbons}), a flare ribbon appears at the east side, to the north from and within the east positive parasitic polarity that is connected to the 3D magnetic null point. This implies the magnetic reconnection occurring at the null point. The preliminary analysis of the magnetic field line evolution shows that the reconnection in the null point appears to lead to the formation of a new flux connecting P2 and N2 polarities, which during the later evolution stage becomes the inner erupting MFR. We plan to investigate this scenario in the follow-up study. Then the ribbons continue separating and elongate considerably, reaching and entering polarities P1 and N2. These regions correspond to the footpoints of the outer erupting MFR.

The comparison of the simulated flare ribbons with the observations shows a remarkable agreement in the ribbon spatial structure. In particular, we find agreement in the hook-shaped north ribbon and elongated south ribbon during the initial evolution stage, and the west kink of the south ribbon and its elongation to polarity P1 during the later stage, as well as the flare ribbon appearance in the core of polarity N2 and vicinity of the magnetic null point. The simulation reconnection flux is also in a good agreement with the estimate derived form the observations by \citet{Kazachenko_2017}, namely about $8\times 10^{21}$~Mx in the simulation versus $6\times 10^{21}$~Mx inferred from the observations. The similar flare ribbon properties in the simulation and the observations, the ejections of the plasma density cavity and the magnetic field lines indicate the ongoing flare reconnection process during the evolution of AR~11158. These results suggest an ejective eruption in the simulation rather than the relaxation of the initial magnetic configuration from the MF simulation to a new equilibrium state.

In our simulation, the reconnection flux does not plateau to constant reconnection flux values that we find in the observations (seen in the inset panel in Figure~\ref{fig:ribbons}). This could be due to the simplified thermodynamics in the MHD simulation that does not describe, for instance, the evolution of the flare--accelerated particles and effects of the radiative transfer. Another important discrepancy between our results and observations is the time scale of the development of the flare ribbons. Flare ribbons in the simulation evolve one order of magnitude faster than those observed. Such a fast evolution is caused by the high reconnection rates in the current sheets due to the high numerical diffusivity of the simulation with the limited numerical grid resolution. Nevertheless, our numerical simulation can successfully simulate the structure and topology of the magnetic field of AR~11158, which combined with the photospheric magnetogram lead to the agreement in the morphological evolution of the flare ribbons and the total reconnected flux.

In our hybrid data-driven MF and data-constrained MHD simulations of the AR~11158 evolution, the initial magnetic configuration obtained from the magnetofrictional simulation appears to be already out of equilibrium. We therefore do not analyse if the conditions for the development of the ideal MHD (torus or kink) instabilities are satisfied \citep{Torok_etal_2004A&A...413L..27T, Kliem_Torok_2006PhRvL..96y5002K}. In the follow-up study, we plan to examine several magnetofrictional initial states to study the transition from stable equilibrium to non-equilibrium MHD evolutions and find the evidence in favour of one of the instabilities mentioned above.

\section{Conclusions} \label{sec:concluions}

We presented our first results of the hybrid \dd MF and data--constrained MHD simulations of AR~11158. To our knowledge, the results of this hybrid approach were reported for the first time. We applied the MF approach to build the coronal magnetic configuration corresponding to the observed SDO/HMI photospheric magnetograms. We used the JSOC PDFI\_SS electric field inversions to drive the bottom boundary of the simulation domain. As a result, we obtained a pre-eruptive non-force-free magnetic configuration containing twisted magnetic fluxes and a 3D magnetic null point. One of those twisted fluxes, detected above the collisional PIL corresponded to the MFR identified in the previous studies and seen in the EUV observations as a sigmoid. We then used the pre-eruptive MF state at about 1.5~hour before the observed X-class flare as the initial state for the MHD simulation, assuming a stratified polytropic solar corona.

The MHD run demonstrated that the initial magnetic configuration obtained with the MF approach was out of equilibrium. We found the eruption of a complex magnetic structure consisting of two MFRs. The eruption was seen in the simulation as the ejection of the plasma density cavity out of the simulation domain. Moreover, we found the ejection of the magnetic field lines related to the density cavity, as well as the development of the system of flare ribbons in good agreement with the observations. To detect the flare ribbons in the simulation, we analysed the plasma temperature enhancements propagating along the magnetic field lines from the reconnection site towards their footpoints. That technique allowed us to capture the main morphological features of the observed flare ribbons in that eruptive event.

The main advantage of our hybrid approach is that it requires significantly less computational resources than the fully \dd full-MHD simulations throughout the course of the flux emergence of AR 11158. The 5-day long formation of the pre-eruptive magnetic configuration in the solar corona was calculated using the MF approach, and the MHD run included only 30~min of the eruptive evolution of AR~11158. Moreover, compared to the previously used hybrid models based on the NLFFF or potential magnetic field, the MF approach allowed our model to incorporate the stored magnetic stresses in the AR, which resulted in the eruption during the MHD run. We therefore conclude that the combination of the \dd MF and data--constrained MHD simulations is a useful practical tool for understanding the 3D magnetic structures of real solar ARs that are unobservable otherwise.

\begin{acknowledgments}

We thank the anonymous referee for helpful comments that improved the paper. We thank the SDO/HMI and SDO/AIA teams for providing us with data. We thank US taxpayers for providing the funding that made this research possible. We acknowledge support of NASA ECIP NNH18ZDA001N and NASA LWS NNH17ZDA001N (A.N.A. and M.D.K.), NASA LWS 80NSSC19K0070 (A.N.A. and Y.F.), DKIST Ambassador Program (A.N.A.). Funding for the DKIST Ambassadors program is provided by the National Solar Observatory, a facility of the National Science Foundation, operated under Cooperative Support Agreement number AST-1400450. Y.F.'s work is also supported by the National Center for Atmospheric Research (NCAR), which is a major facility sponsored by the National Science Foundation under Cooperative Agreement No. 1852977. The results were discussed at the ISSI-Bern meeting ``Data-driven 3D Modeling of Evolving and Eruptive Solar Active Region Coronae''. Resources supporting this work were provided by the NASA High-End Computing (HEC) Program through the NASA Advanced Supercomputing (NAS) Division at Ames Research Center.

\end{acknowledgments}
\bibliographystyle{aasjournal}
\bibliography{arxiv-version}
\end{document}